\documentclass{article}
\usepackage{spconf,amsmath,amssymb,amsfonts,graphicx,booktabs,multirow,array,tabularx,textcomp,algorithmic,cite,hyperref,cleveref}
\usepackage[table]{xcolor}
\usepackage{orcidlink}
\def\BibTeX{{\rm B\kern-.05em{\sc i\kern-.025em b}\kern-.08em
    T\kern-.1667em\lower.7ex\hbox{E}\kern-.125emX}}

\title{Mitigating Accent--Language Confusion in Self-Supervised Speech Representations for Language Identification}
\name{Minu Kim\orcidlink{0009-0005-4520-5071}$^{1}$, Jihwan Lee\orcidlink{0009-0009-3033-2042}$^{1}$, David R. Mortensen\orcidlink{0000-0002-3927-6851}$^{2}$, Shrikanth Narayanan\orcidlink{0000-0002-1052-6204}$^{1}$}%\vspace{-10pt}}
\address{$^{1}$Signal Analysis and Interpretation Lab (SAIL), University of Southern California, USA\\
$^{2}$Language Technologies Institute, Carnegie Mellon University, USA\\
\small\texttt{minukim@usc.edu}}
\begin{document}
%\ninept
%
\maketitle
\begin{abstract}
Spoken language identification (LID) aims to recognize the target language regardless of accent. In practice, however, LID models fine-tuned from self-supervised speech representations frequently confuse accents with languages, misclassifying non-native (L2) speech as the speaker's first language (L1). We show that non-native speech representations lie between native target-language and native L1 poles, causing systematic misclassification. To address this, we introduce a geometric projection that estimates an L1-bias direction solely from native speech and removes it before the frozen LID head. Across five MMS-LID models and non-native corpora, this projection substantially improves target language identification for L2-accented speech while preserving predictions for native speech. These results show that accent-induced L1 bias can be corrected directly within the representation space without L2 training data or model adaptation.
\end{abstract}


\section{Introduction}
\label{sec:introduction}

What should a language identification (LID) system output for non-native (L2) accented speech: accent or language?
% when speech is produced in one language but carries the accent of another? 
It should recognize the language being spoken rather than the accent origin. However, its prediction accuracy degrades with accent strength~\cite{kukk2022improving}, with errors often shifting toward the speaker's first language (L1)~\cite{bafna2025lid}, leading to \textit{accent--language confusion}. This confusion derives from interlanguage phonology transfer~\cite{major2001foreign}, where L2 speakers often realize unfamiliar L2 sounds through L1's phonological categories~\cite{park2025compositional,tepperman2009articulatory}, pulling LID predictions toward the L1. LID systems thus identify a speaker's linguistic background rather than the language being spoken.

This issue is particularly evident in LID systems built on self-supervised speech models (S3Ms)~\cite{fan2021exploring,tjandra2022improved,lee2025l1}. To mitigate such L1 bias, existing approaches combine acoustic predictions with lexicon-free automatic speech recognition (ASR) outputs~\cite{kukk2022improving}, aggregate short chunks, incorporate phonetic sequences~\cite{bafna2025lid}, or fine-tune models on code-switched speech~\cite{patra2026improving}. However, these methods require extra processing, components, or parameter adaptation.

In this work, we propose a method to mitigate the accent--language confusion directly within the representation geometry without any additional data or training.
We draw on prior work showing that phonetic information in S3Ms is organized along linear directions~\cite{mohamed2024orthogonality,choi2026self} and that language bias can be removed via residualization~\cite{kim2026language,kim2026orthogonalization}. Building on this, we estimate the L1 bias in non-native speech by fitting a linear separator solely between native speech in the target language and native speech in the speakers' first languages (L1s). We then project out this rank-one component before classification.

We investigate five MMS-LID models~\cite{pratap2024scaling} on three L2-accented English corpora (Svarah~\cite{javed2023svarah}; L2-ARCTIC~\cite{zhao2018l2}; ALLSSTAR~\cite{bradlow2011language}) spanning diverse L1 backgrounds. Across these datasets, misclassified English utterances disproportionately shift toward the speaker's L1 or a language genealogically related to or in contact with it. This pattern is consistent with contact-driven phonetic convergence~\cite{kim2025improving} and the language relationships captured by MMS-LID~\cite{kim2026scaling}. Without using L2-accented speech for direction estimation, our projection selectively restores English identification under strong L1 bias, whereas removing a random direction yields no such gain. Crucially, predictions for native speech and correctly classified L2 utterances remain fully intact.

Our contributions are threefold: (1) We demonstrate that L2-accented English errors in MMS-LID reflect the speaker's L1 and its related languages. (2) We propose a closed-form projection to eliminate the L1-biased direction using solely native reference speech, requiring no retraining or non-native data for axis estimation. (3) Across five MMS-LID models and three corpora, our approach selectively restores misclassified non-native English utterances while leaving native and correctly classified L2 speech fully intact.

\section{Method}
\label{sec:method}

We propose a geometric projection method to remove L1 bias directly within the representation space without retraining. Let \(a\) be a speech utterance and
\begin{equation}
    x=f_{\theta}(a)\in\mathbb{R}^{d},
    \qquad
    \hat{y}=g_{\phi}(x),
\end{equation}
where \(f_{\theta}\) is a pretrained speech encoder, \(d\) is the representation dimension, and \(g_{\phi}\) is its original LID head. For non-native English, \(g_{\phi}\) can predict the speaker's first language (L1) rather than English. Our goal is to correct this behavior without updating either \(\theta\) or \(\phi\).

\begin{figure}[t]
    \centering
    \includegraphics[width=\columnwidth]{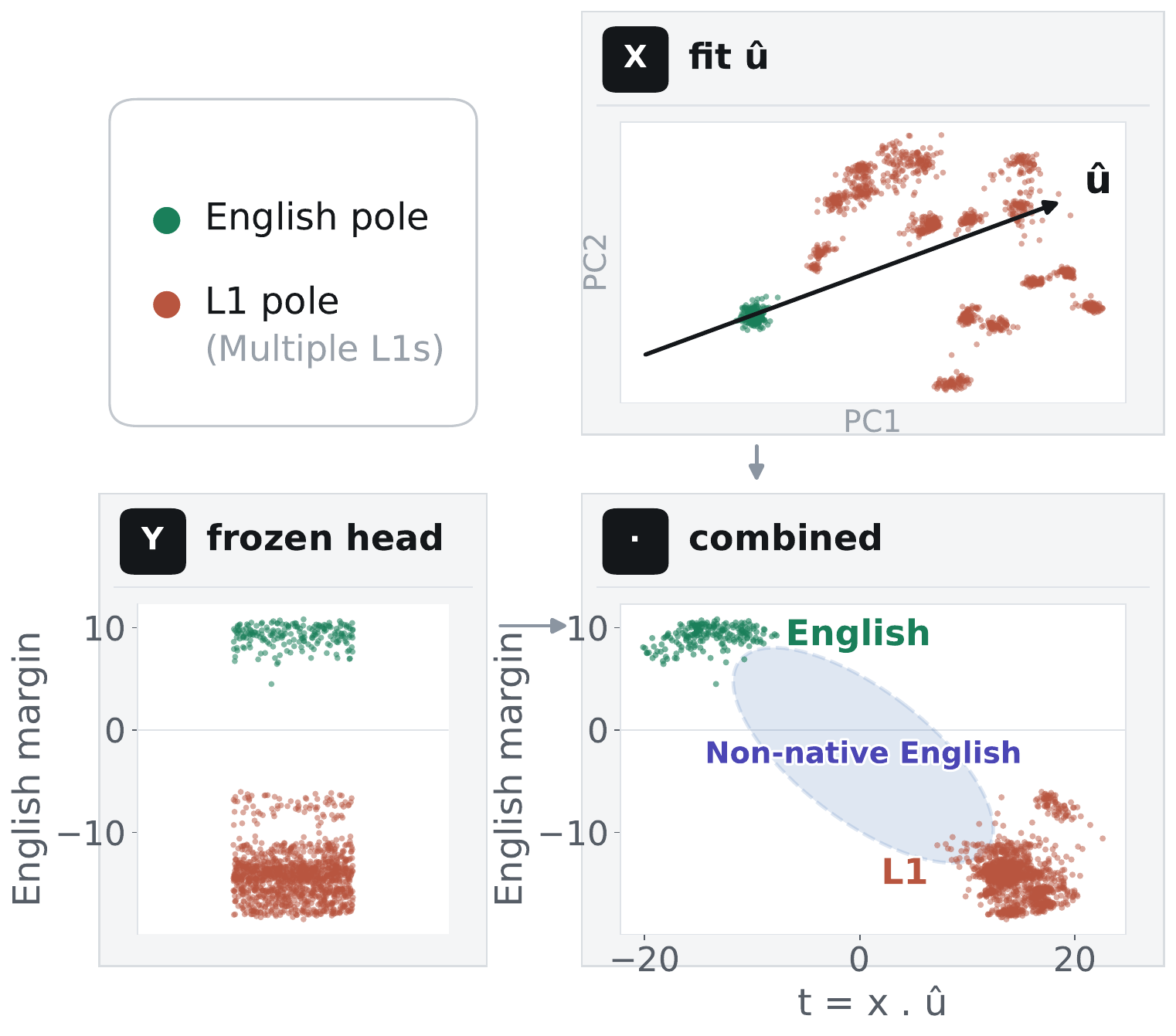}
    \caption{\textbf{Geometry of L1-biased representations illustrated for native and non-native English speech.} 
Top-left (Poles): Native reference distributions for English and non-English L1s. 
Top-right (X): Axis $\hat{u}$ estimated from native speech (English+non-English). 
Bottom-left (Y): Decision margin from the frozen LID head. 
Bottom-right (Combined): Projections ($t = x \cdot \hat{u}$) show L2-accented English positioned between native English and L1s, causing L1-related misclassifications.}
    \label{fig:method_geometry}
\end{figure}

\subsection{Native-only direction estimation}

Our key hypothesis is that non-native English features are displaced along the geometric axis separating native English and native L1 speech. Consequently, this L1-biased direction can be estimated entirely without non-native samples. As conceptualized in Figure~\ref{fig:method_geometry}, evaluating the decision margin along this estimated axis illustrates how non-native speech can occupy an intermediate space.

We construct a native reference set
\begin{equation}
    \mathcal{D}_{\mathrm{nat}} 
    = 
    \{(x_i, z_i)\}_{i=1}^{N},
\end{equation}
containing a total of $N$ reference representations $x_i=f_{\theta}(a_i)$, where $z_i=0$ denotes native English and $z_i=1$ denotes native speech in non-English L1s. No non-native (L2) speech is included.

We estimate the native English--L1 axis using an $\ell_2$-regularized linear separator:
\begin{equation}
\begin{aligned}
    (w^*,b^*)
    =
    \arg&\min_{w,b}\quad
    \frac{1}{2}\lVert w\rVert_2^2 \\
    &+ \sum_{(x_i,z_i)\in\mathcal{D}_{\mathrm{nat}}}
    \mathcal{L}_{\mathrm{CE}}
    \left(
        z_i,\,
        \sigma(w^\top x_i+b)
    \right),
\end{aligned}
\end{equation}
where \(\sigma(t)=(1+e^{-t})^{-1}\). The normalized classifier weight
\begin{equation}
    \hat{u}
    =
    \frac{w^*}{\lVert w^*\rVert_2}
\label{eq:direction_estimation}
\end{equation}
defines the L1-biased direction. Once $\hat{u}$ is estimated, this auxiliary separator is discarded and no longer used.

\begin{table}[t!]
\centering
\caption{\textbf{Per-L1 clip counts in the evaluation corpora.} Only non-native English clips from speakers whose L1 is represented in \textsc{Fleurs} are retained.}
\label{tab:l1_counts}
\small
\setlength{\tabcolsep}{3pt}
\begin{tabular}{@{}lr lr lr@{}}
\toprule
\multicolumn{2}{c}{Svarah}
& \multicolumn{2}{c}{L2-ARCTIC}
& \multicolumn{2}{c}{ALLSSTAR} \\
\cmidrule(r){1-2}
\cmidrule(lr){3-4}
\cmidrule(l){5-6}
Nepali     & 473 & Spanish    & 600 & Cantonese  & 551 \\
Kannada    & 347 & Hindi      & 450 & Mandarin   & 458 \\
Malayalam  & 311 & Mandarin   & 450 & Turkish    & 382 \\
Urdu       & 292 & Korean     & 300 & Spanish    & 377 \\
Odia       & 276 & Vietnamese & 300 & Korean     & 370 \\
Telugu     & 214 & Arabic     & 150 & Portuguese & 179 \\
Tamil      & 200 &            &     & Russian    & 143 \\
Hindi      & 173 &            &     & Hindi      & 140 \\
Gujarati   & 164 &            &     & Vietnamese & 136 \\
Bengali    & 141 &            &     & Hebrew     & 124 \\
Marathi    & 138 &            &     & Japanese   & 102 \\
Assamese   & 111 &            &     & Farsi      &  84 \\
Punjabi    & 111 &            &     & German     &  54 \\
Sindhi     &  84 &            &     & French     &  31 \\
           &     &            &     & Gujarati   &  30 \\
           &     &            &     & Greek      &  27 \\
           &     &            &     & Indonesian &  26 \\
\cmidrule(r){1-2}
\cmidrule(lr){3-4}
\cmidrule(l){5-6}
Total & 3{,}035 & Total & 2{,}250 & Total & 3{,}214 \\
\bottomrule
\end{tabular}
\end{table}

\subsection{L1 bias removal via geometric projection}

To remove L1 bias, we project each representation vector $x$ away from the estimated direction $\hat{u}$:
\begin{equation}
    x'
    =
    \left(I-\hat{u}\hat{u}^{\top}\right)x
    =
    x-\left(\hat{u}^{\top}x\right)\hat{u},
    \label{eq:projection}
\end{equation}
where \(I\in\mathbb{R}^{d\times d}\) is the identity matrix. The scalar \(\hat{u}^{\top}x\) is the coordinate (or magnitude) of \(x\) along the L1-biased direction. As \(\hat{u}\hat{u}^{\top}\) has rank one, Eq.~\eqref{eq:projection} removes only this one-dimensional component.

The corrected representation is passed directly to the frozen LID head:
\[
\boxed{
x
\;\xrightarrow{\;\text{project out }\hat{u}\;}
x'=
\left(I-\hat{u}\hat{u}^{\top}\right)x
\;\xrightarrow{\;\text{frozen LID head }g_{\phi}\;}
\hat{y}'
}
\]
Thus, the correction requires no L2-accented speech and no model retraining.

\section{Experimental Setup}
\label{sec:setup}

\subsection{Data}
\label{sec:data}

We conduct experiments on three English corpora spanning diverse speaker L1 backgrounds: Svarah~\cite{javed2023svarah}, L2-ARCTIC~\cite{zhao2018l2}, and ALLSSTAR~\cite{bradlow2011language}. We retain utterances whose speaker L1 is represented in \textsc{Fleurs}~\cite{conneau2023fleurs}, yielding 3{,}035, 2{,}250, and 3{,}214 clips from 14, 6, and 17 L1 groups, respectively (Table~\ref{tab:l1_counts}). We use native English and the corresponding native L1 speech from the \textsc{Fleurs} dev split to estimate the direction in Eq.~\eqref{eq:direction_estimation}. Its effect on native-language predictions is evaluated separately on the \textsc{Fleurs} test split. No accented-English speech is used for direction estimation.

\subsection{Models and evaluation}

We test five model variants sharing the Massively Multilingual Speech (MMS) backbone~\cite{pratap2024scaling}: MMS-LID-256\footnote{\url{https://huggingface.co/facebook/mms-lid-256}}, MMS-LID-512\footnote{\url{https://huggingface.co/facebook/mms-lid-512}}, MMS-LID-1024\footnote{\url{https://huggingface.co/facebook/mms-lid-1024}}, MMS-LID-2048\footnote{\url{https://huggingface.co/facebook/mms-lid-2048}}, and MMS-LID-4017\footnote{\url{https://huggingface.co/facebook/mms-lid-4017}}. For each model, we extract its representation $x$ from the last encoder layer before the classifier, apply Eq.~\eqref{eq:projection}, and pass $x'$ to the original frozen LID head. We evaluate the English identification rate (\% of clips predicted as English) on native English, non-native English, and native L1 (non-English) speech. To validate that the estimated axis captures meaningful L1 directions, we compare our approach against 20 random control directions.

\vspace{-5pt}
\begin{figure}[t]
    \centering
    \includegraphics[width=\columnwidth]{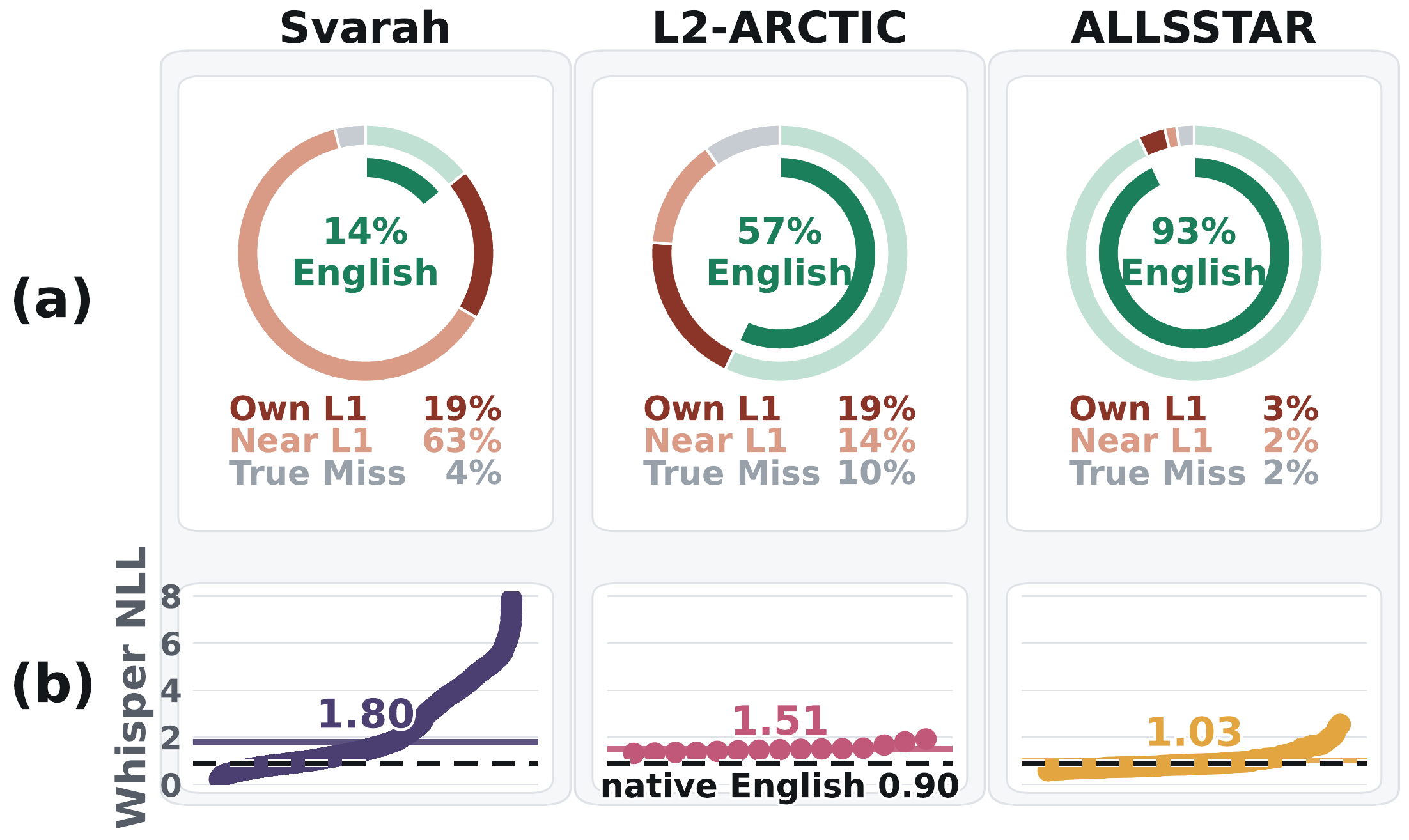}
    \caption{\textbf{Accent-induced confusion in MMS-LID-4017.}
    (a) Prediction breakdown for non-native English speech, categorized into English, speaker's own L1 (\emph{Own L1}), languages related to or in contact with L1 (\emph{Near L1}), and remaining languages (\emph{True Miss}).
    (b) Mean Whisper token-normalized negative log-likelihood (NLL) per speaker (ordered by NLL), with horizontal lines indicating corpus averages and dashed lines denoting the \textsc{Fleurs} native English baseline.}
    \label{fig:confusion}
\end{figure}

\begin{figure*}[t!]
    \centering
    \includegraphics[width=\textwidth]{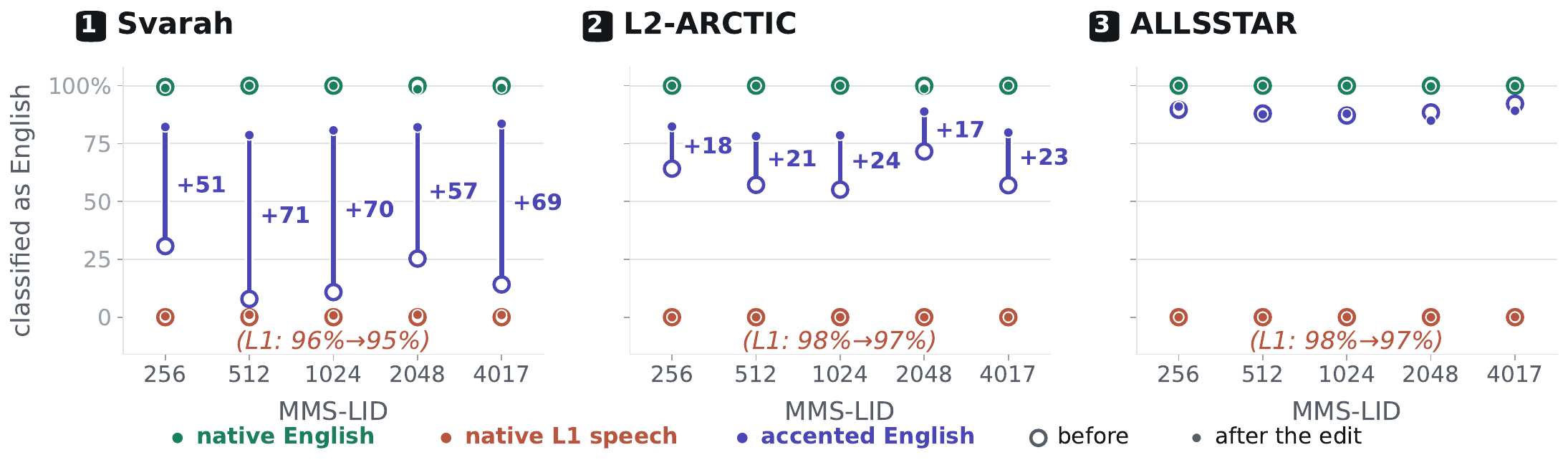}
    \caption{\textbf{English identification performance across models and corpora.} Open and filled markers show English identification rates before and after projection, respectively. Blue numbers indicate percentage-point changes for non-native English, while red annotations at the bottom show model-averaged native L1 identification rates before and after projections. The projection substantially improves English identification on Svarah and L2-ARCTIC while preserving predictions for both native groups.}
    \label{fig:main_results}
\end{figure*}

\begin{figure*}[t!]
    \centering
    \includegraphics[width=\textwidth]{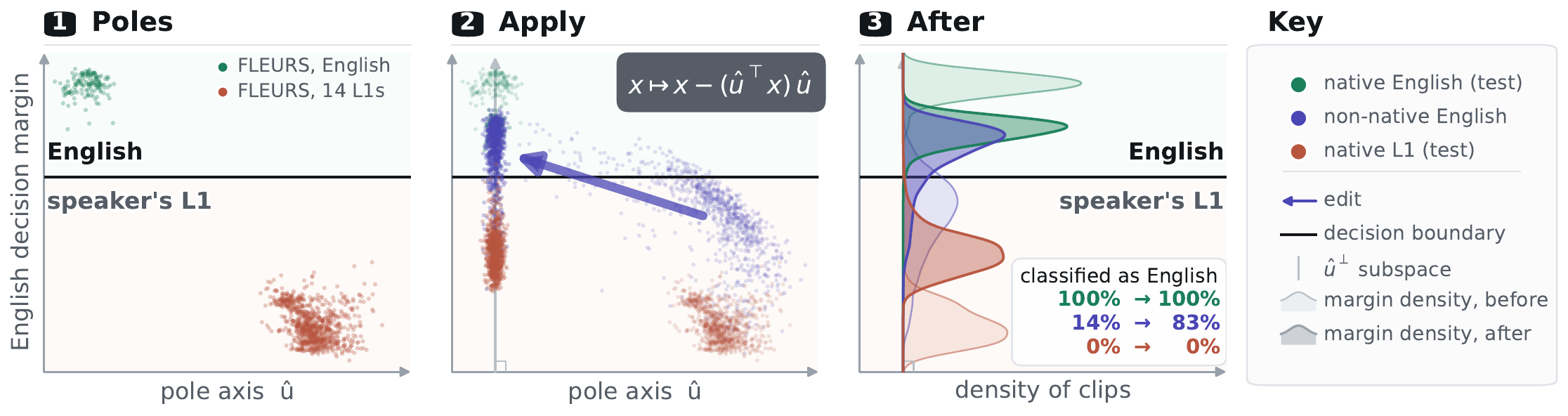}
    \caption{\textbf{Effect of removing the native English--L1 direction.}
    For MMS-LID-4017, \(\hat{u}\) is estimated solely from the \textsc{Fleurs} dev set and visualized on test representations from \textsc{Fleurs} and Svarah. Projecting \(\hat{u}\) out moves non-native English representations toward the English side of the frozen LID decision boundary while preserving representations for both native groups.}
    \label{fig:projection_effect}
    %\vspace{-1mm}
\end{figure*}

\section{Results}
\label{sec:results}

\subsection{L1-dependent accent-language confusion}
\label{sec:accent_confusion}

We examine the misclassification patterns of LID models on non-native speech. Figure~\ref{fig:confusion}(a) shows that MMS-LID-4017 predicts 14\%, 57\%, and 93\% of non-native English clips in Svarah, L2-ARCTIC, and ALLSSTAR as English, respectively. Most misclassifications fall into the speaker's own L1 or a language
genealogically related to or in contact with it (e.g., Hindi-accented English $\rightarrow$ Punjabi/Telugu; Spanish-accented English $\rightarrow$ Catalan; Cantonese-accented English $\rightarrow$ Mandarin). This structured pattern is consistent with contact-driven phonetic convergence~\cite{kim2025improving} and the language relationships captured by MMS-LID~\cite{kim2026scaling}. Similar trends hold across all model scales (\Cref{sec:cross_model_results}).

Following prior work that estimates accentedness from posterior mismatch in automatic speech recognition (ASR)~\cite{rasipuram2015automatic}, we use Whisper~\cite{radford2023robust}'s token-normalized negative log-likelihood (NLL) as a proxy for accent strength. Figure~\ref{fig:confusion}(b) shows that mean NLL decreases from 1.80 on Svarah to 1.51 on L2-ARCTIC and 1.03 on ALLSSTAR (versus 0.90 for native English on \textsc{Fleurs}), inversely mirroring the English identification rate. Thus, greater accent strength corresponds to higher L1-biased confusion.

\subsection{Results across models and corpora}
\label{sec:cross_model_results}

Figure~\ref{fig:main_results} shows that the proposed projection generalizes across all MMS-LID variants. English identification rates for non-native English improve by 51--71\%p on Svarah and 17--24\%p on L2-ARCTIC, whereas the effect is minimal on ALLSSTAR where non-native English is already correctly identified. This corpus-level pattern reflects baseline L2-accent strength (\Cref{sec:accent_confusion}); indeed, the performance gain across L1 groups strongly correlates with Whisper NLL ($r = +0.77$). Meanwhile, identification rates for native English and native L1 speech change negligibly, confirming that the projection corrects L1-biased misclassifications without indiscriminately shifting speech toward English.

To examine this mechanism, we visualize MMS-LID-4017 on Svarah in Figure~\ref{fig:projection_effect}. The direction $\hat{u}$, estimated solely from the \textsc{Fleurs} dev set, separates the two native poles on test data. As non-native English representations lie between these poles, they are often misclassified as L1 or related languages to L1. Removing this single component redirects most non-native English representations to English while keeping native English and native L1 (non-English) speech on their respective sides, isolating accent-induced L1 bias without collapsing the native English--L1 distinction.

\begin{figure}[t!]
    \centering
    \includegraphics[width=\columnwidth]{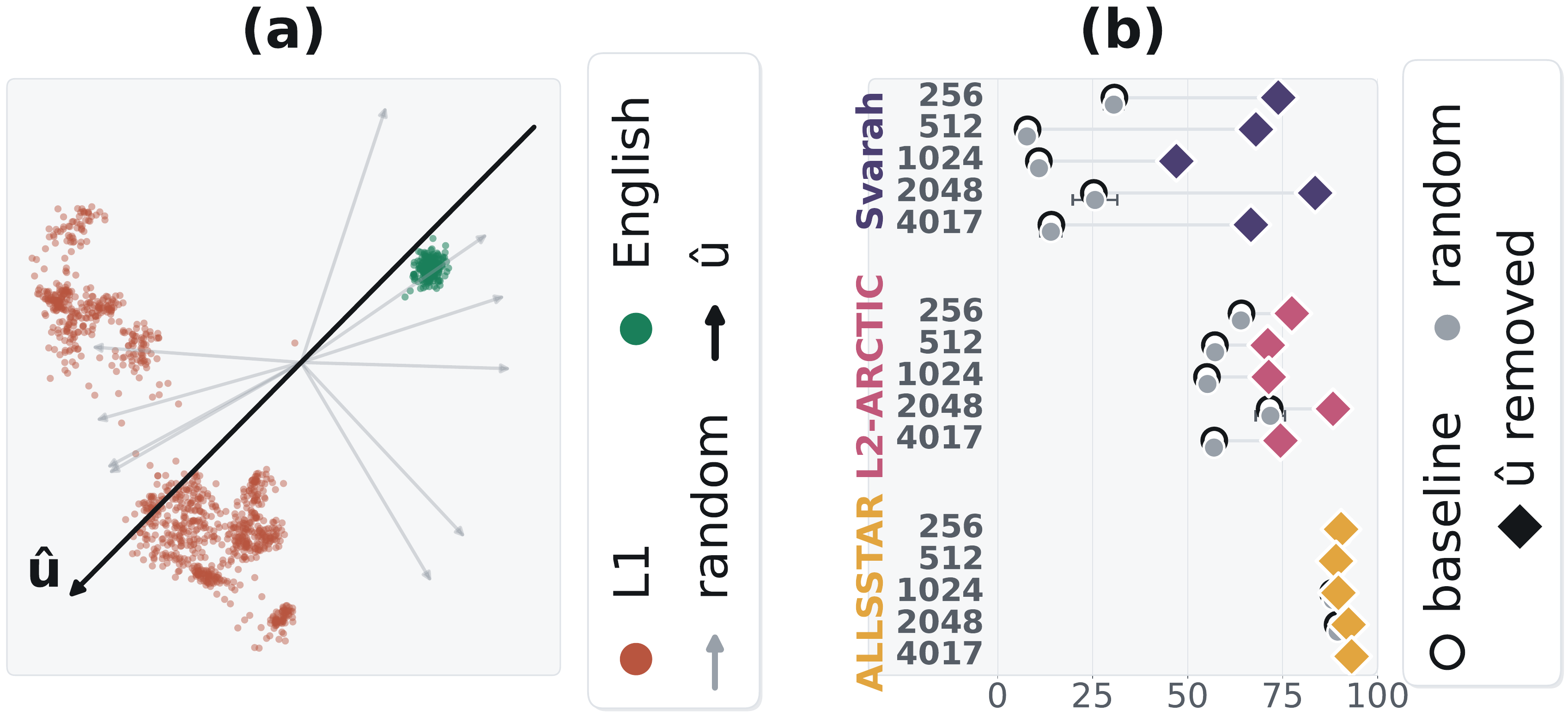}
    \caption{\textbf{Validity of the L1-bias direction.}
    (a) Alignment of \(\hat{u}\) versus 20 random directions with the native English--L1 axis (MMS-LID-4017).
    (b) English identification accuracy before projection and after removing \(\hat{u}\) or random directions.}
    \label{fig:random_control}
    \vspace{-5pt}
\end{figure}

\subsection{Validity of the estimated direction}
\label{sec:random_control}

The observed gains depend on isolating the specific L1-bias axis rather than arbitrary subspace modification. Figure~\ref{fig:random_control}(a) shows that $\hat{u}$ aligns with the separation between native English and native L1 speech, unlike random directions. Consequently, removing 20 random control directions yields no improvement over the baseline, whereas removing $\hat{u}$ consistently improves English identification on Svarah and L2-ARCTIC (Figure~\ref{fig:random_control}(b)). This confirms that performance gains arise directly from targeting the identified L1-bias direction.

\section{Conclusion}
\label{sec:conclusion}

We show that language identification errors for non-native English cluster around the speaker's L1 and related languages. We correct this L1 bias by estimating a single native English--L1 direction solely from native speech and projecting it out before the frozen LID head. Across models and corpora, this projection restores L2-accented English identification while leaving native English and native L1 speech unaffected; removing random directions yields no gains. Accent-induced L1 bias can thus be corrected in the representation space without non-native training data or model adaptation.

\vfill\pagebreak

\newpage

\ninept
%\section{Acknowledgment}

\bibliographystyle{IEEEbib}
\begingroup
\bibliography{strings,refs}

\end{document}